\documentclass[conference,twocolumn,a4paper,10pt]{IEEEtran}
\IEEEoverridecommandlockouts

\usepackage{amsmath}
\usepackage{booktabs}
\usepackage{cite}
\usepackage{float}
\usepackage{graphicx}
\usepackage{siunitx}
\usepackage{stfloats}
\usepackage{textcomp}
\usepackage[final]{microtype}
\usepackage{url}
\usepackage{xcolor}

\begin{document}

\title{Energy-Oriented CGLA Mapping of a Memory-Polynomial Digital Predistortion Kernel}

\author{
\IEEEauthorblockN{Takuto Ando and Yasuhiko Nakashima}
\IEEEauthorblockA{Nara Institute of Science and Technology, 8916-5 Takayama-cho, Ikoma, Nara 630-0192, Japan\\
\texttt{ando.takuto.an5@naist.ac.jp}}
}

\maketitle

\begin{abstract}
Memory-polynomial digital predistortion (DPD) evaluates a small fixed coefficient set over a sliding input history, so its reduction step is a complex-MAC workload with local reuse.
We map this DPD reduction kernel onto In-Memory Accelerator eXtension (IMAX), a programmable CPU-Grounded Linear Array (CGLA) composed of a one-dimensional processing-element/local-memory pipeline.
For a $(P,M)=(5,5)$ odd-order memory-polynomial instance, the mapping keeps the 120\,B coefficient set in local memory, advances the five-tap history over 1024-sample tiles, and realizes the 15 order--delay terms as a 33-stage streaming complex-MAC reduction.
The evaluation measures kernel latency and modeled energy.
All measured paths use the same single-precision complex workload of 32 sequences, each with 2048 complex samples, across an IMAX FPGA prototype, a CUDA implementation on an RTX 4090 system, and an ARM-NEON implementation on Jetson AGX Orin.
With this 1024-sample tile configuration, the IMAX FPGA prototype reports 20.201\,ms end-to-end latency and 1.948\,ms kernel-only latency.
Using the previously reported 28\,nm IMAX frequency and power model, the projected IMAX configuration gives 3.14\,ms end-to-end latency and 0.34\,ms kernel-only latency.
The RTX 4090 baseline has the lowest end-to-end latency at 0.484\,ms.
Under model-based platform power accounting and the stated power assumptions, the projected IMAX configuration gives 169.1$\times$ smaller modeled end-to-end energy per batch than the RTX 4090 baseline. This value uses platform power assumptions rather than workload-dependent runtime power or a direct silicon power measurement.
A controlled synthetic PA-model validation checks that the same 15-term form improves test-set NMSE by 26.1\,dB and ACLR by 26.0\,dB.
These results characterize the mapped memory-polynomial DPD reduction on IMAX for the evaluated tile configuration and power model.

\end{abstract}

\begin{IEEEkeywords}
digital predistortion, CPU-Grounded Linear Array (CGLA), programmable ASIC, IMAX, RF front-end
\end{IEEEkeywords}

\section{Introduction}
\label{sec:introduction}

Digital predistortion (DPD) is required in 5G and emerging 6G base-station transmitters to compensate for the nonlinear response of GaN-based power amplifiers driving wideband and Massive-MIMO signals~\cite{ghannouchi2009dpd,dpd_survey}.
DPD improves the linearity--efficiency trade-off of the RF power amplifier, but the digital correction path must stay within the transmitter power budget so that its processing overhead does not offset the RF-side efficiency gain~\cite{ghannouchi2009dpd,dpd_survey}.
The memory-polynomial form~\cite{ding2004memory,morgan2006gmp} repeatedly evaluates the current sample, a short window of delayed samples, and a fixed coefficient set.
The kernel is a streaming complex-MAC workload with unit-stride sample access and a sliding delay-history window over a fixed coefficient set.
We isolate this reduction kernel and map its coefficient reuse, sliding history, and fixed complex-MAC order to a programmable CPU-Grounded Linear Array (CGLA) pipeline.
The evaluated datapath uses single-precision complex samples and coefficients on all measured software and IMAX configurations.

Prior DPD accelerators report platform-specific metrics.
ASIC and FPGA works emphasize integration, power, and RF-side quality for a particular DPD model~\cite{li2025dpdneuralengine,versluis2025sparsedpd}, whereas GPU implementations provide a high-throughput baseline for calibration or batched execution~\cite{dpd_gpu}.
The 15 complex coefficients of a $(P,M)=(5,5)$ memory-polynomial instance occupy only 120\,B and stay resident in CGLA local memory across many tiles.

In the literature reviewed in this paper, we did not find a reported memory-polynomial DPD implementation on a programmable CGLA.
The target is In-Memory Accelerator eXtension (IMAX), a programmable CGLA with a one-dimensional processing-element/local-memory-module (PE/LMM) pipeline, and the fixed coefficient set and the five-tap delay history both fit in a single 16\,KB LMM bank, while input and output samples cross the host/accelerator boundary in order~\cite{imax_access}.

For polynomial order $P$ and memory depth $M$, the output sample $y[n]$ is written as
\begin{equation}
y[n] = \sum_{m=0}^{M-1}\sum_{\substack{p=1\\p\ \mathrm{odd}}}^{P}
a_{p,m}\,x[n-m]\left|x[n-m]\right|^{p-1},
\end{equation}
where $a_{p,m}$ is the complex coefficient for order $p$ and delay $m$.
Following the commonly used odd-order bandpass-equivalent memory-polynomial form for wideband PA correction~\cite{ding2004memory,morgan2006gmp}, we use $p\in\{1,3,5\}$.
Thus, $P=5$ and $M=5$ yield 15 complex order--delay terms per output sample over five memory taps $m\in\{0,\dots,4\}$.
The coefficient set is fixed during kernel execution, and the delay window advances regularly with the input stream.

We implement an IMAX path on the FPGA prototype and compare it with a CUDA implementation on an RTX 4090 host system and an ARM-NEON implementation on Jetson AGX Orin over the same workload.
Prior ASIC, FPGA, and GPU DPD accelerators are used for positioning because their DPD models, RF conditions, precision choices, and reported metrics differ from ours.
Measured FPGA, GPU, and Jetson AGX Orin platforms are reported separately from a 28\,nm IMAX ASIC projection.

This work treats DPD evaluation as a kernel-mapping study rather than a complete transmitter experiment.
The evaluated path assumes fixed coefficients during kernel execution and focuses on the ordered complex-MAC reduction induced by the memory-polynomial basis, rather than a complete real-time DPD transmitter.
Continuous streaming integration, coefficient adaptation, fixed-point quantization, PA-specific measurement, and EVM-constrained deployment are outside the scope of this paper.

The contributions of this paper are as follows.
\begin{itemize}
\item We formulate the odd-order memory-polynomial DPD reduction as a local-reuse complex-MAC kernel suitable for a one-dimensional CGLA pipeline.
\item We present an IMAX mapping for a $(P,M)=(5,5)$ instance, where a 120\,B coefficient set remains LMM-resident and 15 order--delay terms are reduced through a 33-stage PE/LMM pipeline.
\item We separate kernel-only latency from end-to-end latency so that the CGLA reduction can be distinguished from host-side basis expansion, staging, DMA, and prototype boundary overhead.
\item We compare IMAX with CUDA and ARM-NEON baselines on the same single-precision workload and report modeled end-to-end energy from the platform power values in Table~\ref{tab:devices}.
\end{itemize}

\section{Background and Related Work}
\label{sec:overview}

Accelerator-oriented DPD work spans ASIC, FPGA, and GPU implementations~\cite{dpd_survey}.
DPD-NeuralEngine reports a 22\,nm fixed-function ASIC operating at 250\,MSPS for wideband PA linearization~\cite{li2025dpdneuralengine}.
SparseDPD reports a fixed-pipeline FPGA implementation on Zynq-7Z010 with 241\,mW dynamic power for sparse neural DPD~\cite{versluis2025sparsedpd}, and earlier FPGA DPD systems target real-time 5G PA correction with fixed-function logic~\cite{dpd_fpga}.
GPU DPD work targets high-throughput calibration and batched execution rather than a few-watt transmitter-side correction path~\cite{dpd_gpu}.
DPD accelerators must balance model choice, correction latency, RF-side quality, and digital power, and prior studies use different DPD models, bandwidths, linearization objectives, precision choices, and deployment points~\cite{dpd_survey,li2025dpdneuralengine,versluis2025sparsedpd,dpd_gpu}.
These studies therefore serve here as contextual baselines rather than direct numerical comparisons.
The literature reviewed here does not report a memory-polynomial DPD mapping on a programmable CGLA pipeline.
The evaluated kernel uses the commonly used odd-order bandpass-equivalent memory-polynomial basis~\cite{ding2004memory}.
Linearization quality relative to GMP, Volterra, or neural DPD models is not evaluated in this kernel-level accelerator study.
Thus, this paper does not compete directly with fixed-function RF-DPD accelerators on PA linearization quality or real-time RF throughput.
Its focus is the mapping of the memory-polynomial DPD reduction to a programmable CGLA pipeline, where the measured quantities are kernel latency, data movement, and modeled energy under the stated platform assumptions.

\begin{figure}[t]
\centering
\includegraphics[width=\columnwidth]{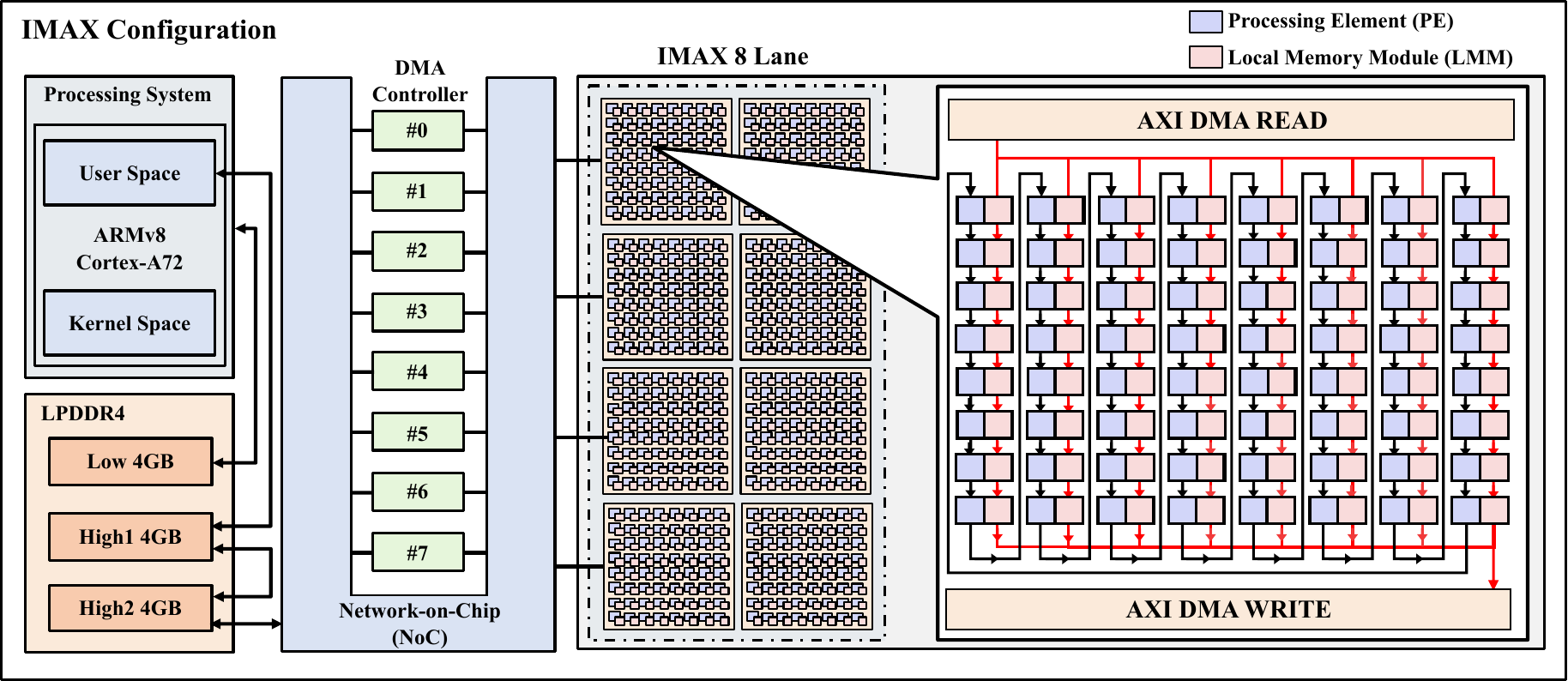}
\caption{IMAX3 FPGA prototype used in this work. The Processing System (PS) hosts the ARM cores and Direct Memory Access (DMA) control, and the Programmable Logic (PL) contains the IMAX array.}
\label{fig:imax3_board}
\end{figure}

IMAX is a programmable CGLA processor with processing elements and local memories arranged along a single pipeline~\cite{imax_access}.
As shown in Fig.~\ref{fig:imax3_board}, the IMAX3 FPGA prototype separates the ARM cores and DMA control on the PS side from the IMAX array on the PL side~\cite{imax_access}.
The host configures the data path on the PS side and launches the stream through DMA toward the PL-side IMAX array~\cite{imax_access}.

Published kernels on IMAX cover FFT and SpGEMM~\cite{imax_access}.
Adjacent spatial-pipeline accelerators, including the Xilinx Versal AI Engine~\cite{gaide2019versal}, the Plasticine reconfigurable fabric~\cite{prabhakar2017plasticine}, and CGRA fabrics such as HyCUBE~\cite{karunaratne2017hycube}, target ML and DSP workloads.
The broader CGRA design space is surveyed in~\cite{cgrasurvey}.
The DPD reduction considered here maps the memory-polynomial kernel onto a programmable CGLA while preserving local coefficient reuse and ordered streaming I/O.
The 15-term coefficient set is small enough (120\,B) to stay LMM-resident across tiles, the sliding history window keeps input DMA traffic compact even after per-pair basis expansion, and the per-output reduction is a one-dimensional dependence chain that places directly onto a linear PE pipeline.
The complex-MAC body uses the IMAX \texttt{exe(OP\_FMA)} and \texttt{exe(OP\_FMS)} pair to handle the sign asymmetry of complex multiplication in two PEs per term~\cite{imax_access}.

\section{IMAX Mapping}
\label{sec:proposed}

\begin{figure*}[t]
\centering
\includegraphics[width=\textwidth]{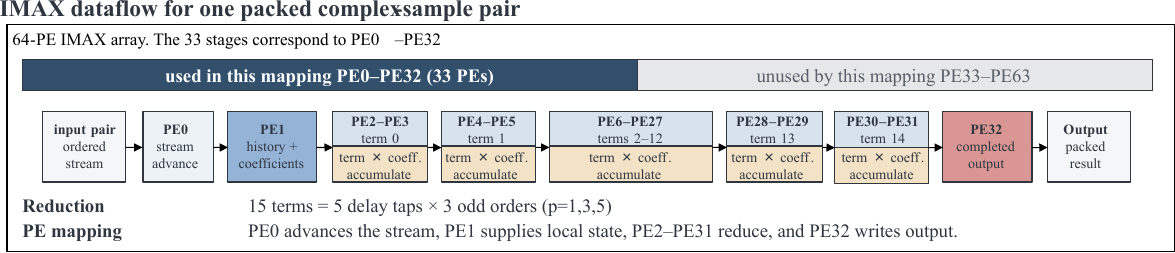}
\caption{IMAX mapping of the complex-MAC reduction for one packed complex-sample pair, where one packed pair contains two adjacent complex samples.
In the MAC stages, a term denotes the pre-expanded order--delay basis $x[n-m]|x[n-m]|^{p-1}$ and a coefficient denotes the resident DPD coefficient $a_{p,m}$.
These terms and coefficients feed a fixed complex-MAC chain, and the final stage writes the accumulated output.
The evaluated $(P,M)=(5,5)$ instance uses a 33-stage segment of the 64-PE array.}
\label{fig:pipeline}
\end{figure*}

Memory-polynomial DPD is mapped to IMAX at the dataflow level.
Fig.~\ref{fig:pipeline} shows the placement used in the evaluated kernel.
Host-side basis expansion forms the order--delay basis terms, and the IMAX pipeline consumes these pre-expanded values while holding coefficients and delay-derived operands in LMM across the tile.
The schedule uses load and store mops for operand movement.
It uses \texttt{OP\_ADD} for offset arithmetic, \texttt{OP\_FML}, \texttt{OP\_FMS}, and \texttt{OP\_FMA} for fused floating-point updates, and \texttt{OP\_NOP} for accumulator forwarding~\cite{imax_access}.

\subsection{Streaming Dataflow}

The mapping separates a tile-invariant coefficient set, a delay history that slides with the stream, and a per-output accumulator.
Input and output samples cross the host/accelerator boundary as ordered streams, while coefficients, delay-derived operands, and the accumulator remain local to the PE/LMM pipeline.
For $(P,M)=(5,5)$, each output sample has 15 complex order--delay terms, and a 1024-sample tile contains 1024 complex samples, i.e., 512 packed complex-sample pairs.
The host-side basis expansion routine \texttt{dpd\_prepare\_tile\_state()} expands the 15 basis terms for the two complex samples in each packed pair into \texttt{term\_re/im}, while the 120\,B coefficient set is initialized once in LMM and reused across the 512 packed pairs.
The per-tile traffic is 122.88\,KB of basis staging plus 8.192\,KB input and 8.192\,KB output sample streams.

Each output sample is computed as one ordered reduction, and no partial sum is written back between polynomial orders or memory delays.
For each $m \in \{0,\dots,4\}$ and odd $p \in \{1,3,5\}$, the mapped IMAX kernel consumes a pre-expanded basis value, multiplies it by $a_{p,m}$, and accumulates the product.
For a term $t=t_{\mathrm{re}}+j t_{\mathrm{im}}$ and coefficient $c=c_{\mathrm{re}}+j c_{\mathrm{im}}$, the update is
\begin{align}
\mathrm{acc}_{\mathrm{re}} &\leftarrow \mathrm{acc}_{\mathrm{re}} + t_{\mathrm{re}}c_{\mathrm{re}} - t_{\mathrm{im}}c_{\mathrm{im}},\\
\mathrm{acc}_{\mathrm{im}} &\leftarrow \mathrm{acc}_{\mathrm{im}} + t_{\mathrm{re}}c_{\mathrm{im}} + t_{\mathrm{im}}c_{\mathrm{re}}.
\end{align}
In the IMAX schedule, term~0 seeds the accumulator with \texttt{exe(OP\_FML)}, and terms~1--14 follow a two-PE pattern in which \texttt{exe(OP\_FMS)} performs the real update and \texttt{exe(OP\_FMA)} performs the imaginary update.
The same instruction pattern repeats over the 15 terms and the 512 packed pairs of a 1024-sample tile.

\subsection{Pipeline Placement}

Each term depends only on the previous accumulator, so IMAX maps the reduction as a one-dimensional spatial chain.
PE0 advances the per-pair basis offset, PE1 issues the seed-term loads, PE2--PE31 host the 15 complex-MAC stages, and PE32 drains and stores \texttt{out\_re}/\texttt{out\_im}.
More generally, an odd-only memory-polynomial $(P,M)$ instance has $T=\lceil P/2\rceil M$ terms and uses $2T+3$ stages, so $(P,M)=(5,5)$ gives $T=15$ and 33 stages.
Fig.~\ref{fig:pipeline} summarizes this placement as a single stream path from local-state setup to writeback.

\section{Evaluation}
\label{sec:experiments}

\subsection{Setup and Power Model}

We evaluate the $P=5$, $M=5$ workload on 32 independent sequences, each with 2048 complex samples, for 65,536 complex samples per batch, and compare the IMAX FPGA/projection configurations with RTX 4090 and Jetson AGX Orin baselines.
Table~\ref{tab:devices} lists the platforms and power values.
\begin{table*}[t]
\centering
\caption{Evaluated platforms and power model. The IMAX configurations use the 33-stage DPD dataflow mapped onto a segment of the 64-PE array.}
\label{tab:devices}
\footnotesize
\resizebox{\textwidth}{!}{%
\begin{tabular}{@{}llll@{}}
\toprule
Platform & Compute path & Configuration & Power model \\
\midrule
IMAX FPGA prototype & measured DPD mapping & 64-PE array, 33-stage segment, 512\,KB LMM, 145\,MHz & --- \\
IMAX ASIC projection & projected DPD mapping & 64-PE array, 33-stage segment, 512\,KB LMM, 840\,MHz, 28\,nm~\cite{imax_access,SynopsysNDDCUltra} & 1.161\,W synthesized ASIC + 0.525\,W ARM host orchestration \\
RTX 4090 host & measured custom CUDA & RTX 4090 + Xeon w5-2465X~\cite{nvidia_ada,Intel_Xeon_w5-2465X_Specs} & 450\,W GPU TGP $+$ 200\,W processor base power \\
Jetson AGX Orin & measured ARM-NEON & Jetson AGX Orin~\cite{nvidia_jetson_agx_orin,nvidia_jetson_orin_power_modes} & 15\,W power profile \\
\bottomrule
\end{tabular}
}
\end{table*}

The FPGA, RTX 4090, and Jetson AGX Orin rows are measured, whereas the IMAX ASIC row is a 28\,nm projection.
Unless stated otherwise, the representative IMAX row uses a 1024-sample tile configuration that amortizes per-tile host, register, and transfer overhead while preserving the same 65,536-complex-sample batch workload.
With this configuration, the FPGA prototype reports a 10-run average of 20.201\,ms end-to-end latency and 1.948\,ms kernel-only latency for the 33-stage DPD reduction.
End-to-end time includes host-side basis expansion, staging/DMA, register setup, IMAX execution, and output writeback, whereas kernel-only latency refers only to the IMAX complex-MAC reduction over pre-expanded basis values.
Thus, host-side basis expansion is included in the reported end-to-end timing, while moving basis generation into the array or a streaming front-end is left outside this kernel-mapping evaluation.
All measured paths use the same single-precision complex workload, where each complex sample and coefficient is represented as two 32-bit floating-point values.
Using the 145\,MHz FPGA prototype point and the 840\,MHz 28\,nm point from prior IMAX synthesis results~\cite{imax_access}, reported under a Synopsys Design Compiler flow with a TSMC 28\,nm library~\cite{SynopsysNDDCUltra}, the measured 1.948\,ms kernel-only time is scaled to 0.336\,ms and combined with 2.3\,ms ARM preparation and 0.5\,ms on-chip memory access to give 3.14\,ms end-to-end.

\subsection{Latency and Correctness}

\begin{figure}[t]
\centering
\includegraphics[width=\columnwidth]{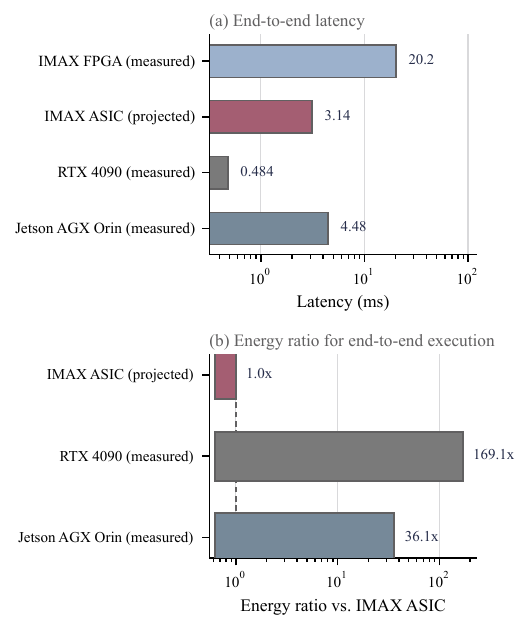}
\caption{Latency and modeled energy per batch for the evaluated batch-32 workload.
Top: end-to-end latency, where the RTX 4090 is measured and the IMAX ASIC row is projected.
Bottom: end-to-end energy per batch, normalized to the projected IMAX ASIC configuration under the power model in Table~\ref{tab:devices}; it is not a workload-dependent runtime power measurement.}
\label{fig:platform_results}
\end{figure}

Fig.~\ref{fig:platform_results}(a) reports end-to-end latency.
RTX 4090 is shortest at 0.484\,ms, the projected IMAX ASIC configuration takes 3.14\,ms, Jetson AGX Orin takes 4.478\,ms, and the FPGA prototype takes 20.201\,ms including host boundary, register setup, and tile transfers in the 1024-sample tile configuration.
The 1.948\,ms FPGA kernel-only measurement isolates the PE/LMM mapping from prototype boundary overhead.
For the FPGA prototype, the difference between 20.201\,ms end-to-end latency and 1.948\,ms kernel-only latency indicates that most of the observed time is outside the PE/LMM reduction, in prototype host-boundary and DMA setup.
The FPGA, GPU, and Jetson AGX Orin platforms are measured against the same reference output, and only the IMAX 28\,nm configuration is projected.
Maximum absolute output differences remain in the $10^{-6}$ range for IMAX and the $10^{-7}$ range for the GPU and Jetson platforms, with the IMAX difference coming from FMA/FMS accumulation order.

A 32-sample tile gives a small-tile streaming-oriented reference point. It produced no mismatches in a 10-run check, with 66.984\,ms end-to-end and 6.034\,ms kernel-only latency.
At the other end, a 2048-sample tile produced no mismatches and reached 19.740\,ms end-to-end and 2.905\,ms kernel-only latency, but it buffers a full block before the IMAX call.
We therefore treat tile size as a dataflow parameter. The 1024-sample row is the representative overhead-amortized operating point, while the 2048-sample row is only an auxiliary full-block check.

\subsection{Energy per Batch}

The energy comparison is scenario-based accounting under the stated platform power assumptions.
It uses the platform power values in Table~\ref{tab:devices}, rather than workload-dependent runtime power measurements, and is not a direct cross-platform silicon power measurement.
Energy is computed as average power times wall time.
\begin{equation}
E_{\mathrm{batch}} = P_{\mathrm{model}} \times T_{\mathrm{wall}}.
\end{equation}
The RTX 4090 platform uses 450\,W GPU TGP plus 200\,W Xeon processor base power~\cite{nvidia_ada,Intel_Xeon_w5-2465X_Specs}, the Jetson platform uses its 15\,W power profile~\cite{nvidia_jetson_orin_power_modes}, and IMAX uses 1.161\,W for the 0.336\,ms kernel plus 0.525\,W for 2.80\,ms of ARM host orchestration~\cite{imax_access}.
The projected IMAX ASIC configuration consumes 1.86\,mJ per batch, while RTX 4090 consumes 314.6\,mJ ($169.1\times$) and Jetson AGX Orin consumes 67.2\,mJ ($36.1\times$).
This corresponds to 20.9\,MS/s, 313\,M complex-MAC/s, and 28.4\,nJ/sample end-to-end for the projected IMAX row. Its kernel-only portion consumes 0.39\,mJ, or 6.0\,nJ/sample.

PA-specific RF measurement, fixed-point quantization, continuous streaming, and coefficient adaptation remain outside this accelerator evaluation.

\subsection{Synthetic PA Validation}

For a controlled synthetic PA-model validation, we apply the same 15-term correction form to a synthetic $(P,M)=(5,5)$ memory-polynomial PA model with a 65,536-sample OFDM-like waveform.
Metrics are computed only on a held-out test waveform.
The correction improves NMSE from -18.75\,dB to -44.87\,dB and ACLR from -31.26\,dBc to -57.24\,dBc, while PAPR increases from 10.39\,dB to 13.32\,dB.
This check only confirms that the evaluated 15-term kernel represents a DPD correction workload. Coefficient adaptation and transmitter integration are outside the accelerator evaluation.

\section{Conclusion}
\label{sec:conclusion}

We evaluate a CGLA mapping for the complex-MAC reduction in a $(P,M)=(5,5)$ memory-polynomial DPD kernel.
The mapping keeps the 120\,B coefficient set local in a 33-stage PE/LMM pipeline.
For the representative 1024-sample tile, the FPGA prototype records 20.201\,ms end-to-end latency and 1.948\,ms kernel-only latency, while the model-based 28\,nm projection gives 3.14\,ms end-to-end and 1.86\,mJ per batch under the stated power values.
The projected energy is a scenario-based estimate, not a workload-dependent runtime power measurement or a direct silicon power measurement.
Future work is coefficient adaptation and continuous-streaming integration around the same kernel mapping.

{\raggedright
\bibliographystyle{IEEEtran}
\bibliography{bibliography_blind}
}

\end{document}